# The Ice Chamber for Astrophysics-Astrochemistry (ICA): A New Experimental Facility for Ion Impact Studies of Astrophysical Ice Analogues

Péter Herczku[1*], Duncan V. Mifsud[2,1*], Sergio Ioppolo[3], Zoltán Juhász[1], Zuzana Kaňuchová[4,5], Sándor T. S. Kovács[1], Alejandra Traspas Muiña[3], Perry A. Hailey[2], István Rajta[1], István Vajda[1], Nigel J. Mason[2], Robert W. McCullough[6], Béla Paripás[7], Béla Sulik[1]

| | |
|---|---|
| 1 | Institute for Nuclear Research (Atomki), Debrecen H-4026, PO Box 51, Hungary |
| 2 | Centre for Astrophysics and Planetary Science, School of Physical Sciences, University of Kent, Canterbury CT2 7NH, United Kingdom |
| 3 | School of Electronic Engineering and Computer Science, Queen Mary University of London, London E1 4NS, United Kingdom |
| 4 | Astronomical Institute, Slovak Academy of Sciences, Tatranská Lomnica SK-059 60, Slovakia |
| 5 | INAF Osservatorio Astronomico di Roma, Monte Porzio Catone, RM-00078, Italy |
| 6 | Department of Physics and Astronomy, School of Mathematics and Physics, Queen's University Belfast, Belfast BT7 1NN, United Kingdom |
| 7 | Department of Physics, Faculty of Mechanical Engineering and Informatics, University of Miskolc, Miskolc H-3515, Hungary |
| * | Corresponding authors: herczku.peter@atomki.hu and duncanvmifsud@gmail.com |

**Author ORCID Identification Numbers**

| | |
|---|---|
| P. Herczku | 0000-0002-1046-1375 |
| D. V. Mifsud | 0000-0002-0379-354X |
| S. Ioppolo | 0000-0002-2271-1781 |
| Z. Juhász | 0000-0003-3612-0437 |
| Z. Kaňuchová | 0000-0001-8845-6202 |
| S. T. S. Kovács | 0000-0001-5332-3901 |
| A. Traspas-Muiña | 0000-0002-4304-2628 |
| P. A. Hailey | 0000-0002-8121-9674 |
| I. Rajta | 0000-0002-5140-2574 |
| I. Vajda | 0000-0001-7116-9442 |
| N. J. Mason | 0000-0002-4468-8324 |
| R. W. McCullough | 0000-0002-4361-8201 |
| B. Paripás | 0000-0003-1453-1606 |
| B. Sulik | 0000-0001-8088-5766 |

**Abstract**

The Ice Chamber for Astrophysics-Astrochemistry (ICA) is a new laboratory end-station located at the Institute for Nuclear Research (Atomki) in Debrecen, Hungary. The ICA has been specifically designed for the study of the physico-chemical properties of astrophysical ice analogues and their chemical evolution when subjected to ionising radiation and thermal processing. The ICA is an ultra-high vacuum compatible chamber containing a series of IR-transparent substrates mounted in a copper holder connected to a closed-cycle cryostat capable of being cooled down to 20 K, itself mounted on a 360° rotation stage and a *z*-linear manipulator. Ices are deposited onto the substrates via background deposition of dosed gases. Ice structure and chemical composition are monitored by means of FTIR absorbance spectroscopy in transmission mode, although use of reflectance mode is possible by using metallic substrates. Pre-prepared ices may be processed in a variety of ways. A 2 MV Tandetron accelerator is capable of delivering a wide variety of high-energy ions into the ICA, which simulates ice processing by cosmic rays, the solar wind, or magnetospheric ions. The ICA is also equipped with an electron gun which may be used for electron impact radiolysis of ices. Thermal processing of both deposited and processed ices may be monitored by means of both FTIR spectroscopy and quadrupole mass spectrometry. In this paper, we provide a detailed description of the ICA set-up, as well as an overview of preliminary results obtained and future plans.

# I. Introduction

The chemistry which occurs in the interstellar medium (ISM) is often significantly different to that which occurs on Earth. Since the discovery of the first molecular species in the 1930s [1], much interest has been paid to the formation mechanisms and fates of interstellar molecules as such molecules may be seeded into planetary bodies during their formation and may be of significant consequence to geological processes as well as the potential development of life. To study the formation and reactivity of these molecules, it is necessary to simulate the interstellar environment in which they are located.

Molecular formation in the ISM takes place largely within dense molecular clouds composed of micron-sized carbonaceous or silicate dust grains. The interiors of such clouds are shielded from the far- and vacuum-UV components of the interstellar radiation field and are characterised by low temperatures (10-20 K) and particle number densities of $10^2$-$10^6$ cm$^{-3}$ [2]. Under such conditions, gas-phase species including atomic and molecular hydrogen, along with traces of atomic carbon, nitrogen, oxygen, other radicals, and volatile molecules (such as CO, NO, and $C_2H_2$) freeze out onto the interstellar dust grains. Accreted species may, having been brought into proximity of one another at the grain surface, react to form an ice mantle enriched in new product molecules including $H_2$, $H_2O$, $NH_3$, $CH_4$, $CO_2$, and $CH_3OH$ [e.g., 3-10]. Laboratory experiments have also provided evidence for the formation of much larger complex organic molecules, including: methyl formate, glycolaldehyde, ketene, and acetaldehyde, among others [e.g., 11-14].

Molecules entrapped within icy grain mantles may also take part in chemistry induced by some form of energetic processing, such as that initiated by X-rays and cosmic rays (i.e., high-energy nuclei), both of which are able to penetrate into the interior of the dense molecular cloud [e.g.,

15-23]. Furthermore, although icy grain mantles within the cloud are shielded from external sources of UV photons, cosmic ray interaction with gas-phase molecular hydrogen is able to produce low-intensity Lyman-α photons which may engender photophysical processes and photochemical reactions [e.g., 2,24-27]. A visual summary of the processes leading to novel chemistry in icy grain mantles is given in Fig. 1.

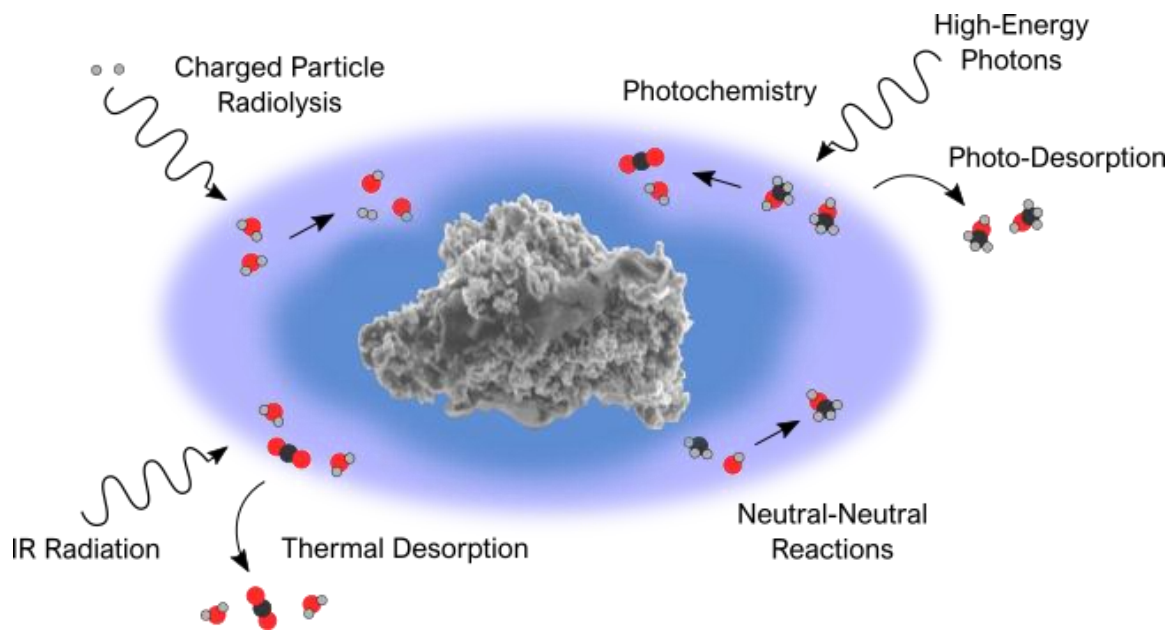

**Fig. 1** Illustrative summary of the chemical processes which may occur in the icy mantles of interstellar dust grains.

The importance of the chemistry occurring within icy grain mantles cannot be underestimated, as it is these reactions which ultimately contribute to the chemical complexity and molecular diversity of stellar, circumstellar, and planetary environments which evolve from such dense molecular clouds. The need to understand such chemical complexity provides a strong motivation for performing laboratory simulations of the energetic processing to which interstellar ices are subjected.

The first apparatus specifically designed for studying the energetic processing of interstellar ice analogues was developed by Greenberg and Allamandola at Leiden Observatory in the Netherlands during the mid-1970s [28]. Their initial apparatus consisted of a high-vacuum chamber within which ice analogues could be formed on a cold (10 K) substrate. Simulation of the Lyman-α photolysis occurring within interstellar clouds was achieved using UV lamps [29] and any resultant structural or chemical changes within the ice were monitored spectroscopically.

Over the subsequent decades, similar experimental set-ups have been established at various laboratories throughout the world. Astrophysical ice analogue processing has not remained limited to Lyman-α photolysis, however, and now includes processing by vacuum- or extreme-UV photons (e.g., by research groups at the Centro de Astrobiología in Spain, the Institut d'Astrophysique Spatiale in France, the National Central University in Taiwan, the ASTRID2 Synchrotron Light Source in Denmark, etc.) and ions (e.g., by research groups at the INAF

Osservatorio Astrofisico di Catania in Italy, the NASA Goddard Space Flight Centre in the United States, the Grand Accélérateur National d'Ions Lourds in France, etc.) [30].

Laboratory astrochemistry experiments have revealed a great deal of information on the chemical reactivity of molecules within interstellar icy grain mantles. However, systematicity, repeatability, and reproducibility have sometimes been difficult to achieve even when using contemporary ultra-high vacuum (UHV) astrochemistry chambers, as ice analogues in different experiments are formed under slightly different deposition conditions [31,32].

In this paper, we present a technical overview of the Ice Chamber for Astrophysics-Astrochemistry (ICA); a new end-station for ion irradiation studies relevant to experimental astrochemistry located at the Institute for Nuclear Research (Atomki) in Debrecen, Hungary. One of the main advantages of the ICA set-up lies in its provision for up to four deposition substrates, which allows for the production of four ice analogue replicates created under identical experimental conditions.

Thus, the ICA is designed to facilitate the performance of systematic studies wherein a selected number of parameters (e.g., ice thickness, morphology, temperature, projectile ion nature and energy, etc.) may be controlled and varied with ease. Additionally, thermal, ion, and electron processing are all readily available at the ICA, allowing for various processing-type combinations. This paper provides a detailed description of the ICA set-up, including the accelerator facilities used to deliver projectile ions. Preliminary results of ion radiolysis experiments carried out under high-vacuum conditions are also presented and discussed in light of planned future work.

## II. The Experimental Facility

### A. Accelerator and Beam Conditions

The use of particle accelerators to deliver energetic projectile ions in laboratory experiments seeking to simulate the effects induced by cosmic ray or solar wind irradiation of astrophysical ices first began in the 1970s and has since grown to become a major branch of research within the field (for a review on the use of accelerators in astrochemistry, see the work by Strazzulla [30]). In this section, we provide a description of the facility used to supply energetic ions to the ICA for ion radiolysis studies of astrophysical ice analogues.

A detailed description of the Atomki Tandetron Laboratory was provided earlier [33]. The nominal terminal voltage of the accelerator is in the range 0.085-2 MV, with a stability of ±200 V hr$^{-1}$. A terminal voltage of 1 MV was found to vary within 50 V over a time range in excess of five hours, with a ripple of 15 V. Such values allow for the production of extremely stable beams of many different ions which may be delivered to any of the nine beamlines currently available at Atomki.

The accelerator is equipped with two multi-cusp ion sources for proton and helium ion beams and a caesium sputtering ion source for heavier ions. The accelerator is able to provide very high beam currents: >200 µA for protons and at least a few µA for heavier ion projectiles (helium through to gold). Although such currents are available, the ion beam currents actually used for ice irradiation experiments typically do not exceed a few hundred nA. For proton beams, the available energy range is 0.2-4 MeV while for heavier ion beams this range is determined by the available terminal voltage range, the relation between which is given by Eq. 1:

$$E_{\text{beam}} = q(V_{\text{ext}} + V_{\text{term}}(n + 1))$$

(Eq. 1)

where $E_{\text{beam}}$ is the projectile ion beam energy, $V_{\text{ext}}$ is the extraction voltage for the particular ion being used, $V_{\text{term}}$ is the terminal voltage (which can be set between 0.085-2 MV), $n$ is the positive integer charge of the projectile ion, and $q$ is the fundamental charge constant ($1.602 \times 10^{-19}$ C) which is taken to be unity when $E_{\text{beam}}$ is expressed in eV. Hence, for example, a $He^+$ ion beam for which the extraction energy is 20 keV has an available energy range of 190-4020 keV.

Such beam energies cover the low-energy side of the Bragg peak, where projectile ions travelling through matter experience their maximum energy transfer which induces efficient molecular fragmentation. Additionally, the use of beams with energies in the keV-MeV range and currents on the nA scale implies a maximum power deposition on the 100 mW order of magnitude, and so macroscopic heating of the ice analogue targets is expected to be insignificant.

The availability of such a wide array of projectile ions at various energies suited to laboratory simulations of astrophysical ice processing is perhaps what makes the ICA set-up unique among the many astrochemistry facilities that exist around the world. Indeed, work conducted at the ICA will complement and extend the findings of studies conducted at other facilities where different projectiles and energy ranges are available, such as the set-up at the INAF Osservatorio Astrofisico di Catania in Italy where ion beam energies of up to 0.4 MeV are available [34], or that at the Grand Accélérateur National d'Ions Lourds in France where ion projectiles at much lower energies (a few tens of keV) or swift heavy projectiles (e.g., Ni or Xe ions) at significantly higher energies (1 MeV per nucleon) are available [35].

A stable and homogeneous current density at the surface of the sample ice analogue is required to accurately monitor physico-chemical changes occurring within the ice via spectroscopy. We have therefore applied an ion beam scanner consisting of $x$ and $y$ deflector pairs, as well as high-voltage ramp generators with minimum frequencies of 25 and 600 Hz in the $x$ and $y$ directions, respectively, so as to provide homogeneous irradiation of the ice target during radiolysis (Fig. 2).

The nominal scanned area at the target is around 25×25 mm. From this area, a circular collimator defines a homogeneous circle with a diameter of 14 mm. The resultant circularly-shaped ion beam can be viewed on a quartz probe using a high-definition camera. During irradiation, the current is continuously monitored by a Faraday cup ($I_{\text{mon}}$) placed in the path of the incident ion beam with a second collimator at its base through which the beam passes towards the target. This second collimator defines the diameter of the beam spot on the sample as 12.6 mm. The true deposited current is measured by a removable reference Faraday cup ($I_{\text{ref}}$), and so knowledge of $I_{\text{mon}}$ and $I_{\text{ref}}$ allows for determination of the actual current density on the irradiated sample. The homogeneity of the beam can be frequently checked during irradiation by using a collimator on the sample holder ($C_2$) and a final Faraday cup ($F_3$), as geometric factors require the measured beam current here ($I_3$) to be half that measured at $I_{ref}$ for all beam currents provided that the ion beam is indeed homogeneous.

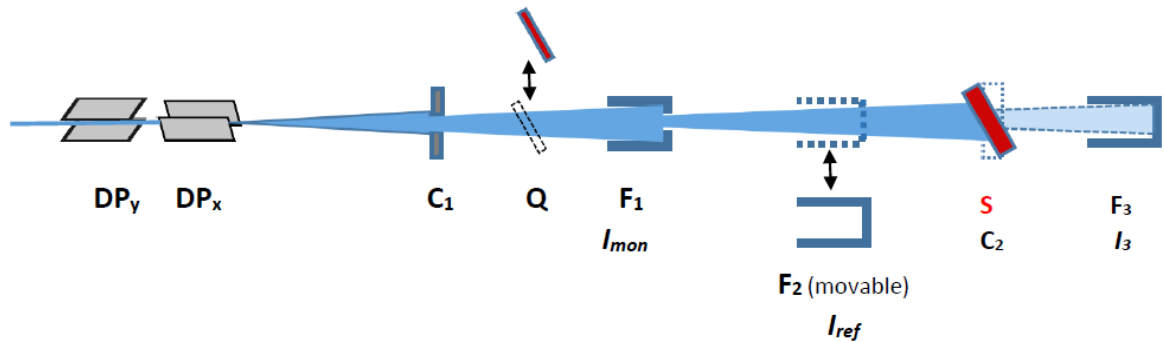

**Fig. 2** Schematic diagram of the controlled ion beam irradiation system (not to scale). Once the ion beam passes through deflector pairs $DP_x$ and $DP_y$, the collimator $C_1$ defines a circular beam with diameter 14 mm which may be measured on the quartz probe Q. The beam current ($I_{mon}$) is measured using the Faraday cup $F_1$ with a second collimator reducing the beam diameter to 12.6 mm at the sample surface. The removable reference cup $F_2$ allows for measurement of the deposited current ($I_{ref}$) just before the beam impacts the irradiated sample S. The beam homogeneity can be confirmed using the collimator $C_2$ and a final Faraday cup $F_3$.

## B. The Ice Chamber: Physical Characteristics and Gas Inlet System

The ICA multi-purpose set-up comprises a spherical decagon UHV-compatible chamber (Kimball Physics) with an inner diameter of 160 mm and having two DN-160 CF and ten DN-40 CF ports. At the centre of this chamber there is a heat-shielded copper sample holder within which there is currently the provision to mount up to four IR-transparent (typically zinc selenide) substrates on which astrophysical ice analogues may be prepared (Fig. 3). The sample holder is moveable along the vertical axis of the chamber and can also be rotated about this axis by means of a $z$-linear manipulator with 50.8 mm stroke and a 360° rotation stage, respectively. This design allows for a maximum of four independent irradiations of ice analogues – created under identical deposition conditions – to be carried out.

The sample holder can be cooled to 20 K using a closed-cycle helium cryostat (Leybold Coolpower 7/25 with a Leybold Coolpak 4000 compressor unit). The temperature across the sample holder is measured using two uncalibrated silicon diodes (Lake Shore DT-670B-CO) connected to a cryogenic temperature controller (Lake Shore model 335), with accurate temperature measurements being made with a proportional integral-differential controller. The actual measured temperatures of the deposition substrates were validated by performing the controlled warmings of pure $N_2O$ and CO ices and comparing the temperatures at which phase changes and ice sublimation we spectroscopically observed to occur with well-established literature values [36,37]. The temperature of the sample holder itself can be regulated by setting an equilibrium between heating and cooling using an internal 25 Ω / 100 W cartridge heater (HTR-25-100). Such an arrangement allows for an effective working temperature range of 20-300 K. Additionally, the sample holder temperature can be varied at a rate of 0.1-10 K min$^{-1}$, allowing certain temperature-programmed desorption (TPD) studies to be performed.

A turbomolecular pump and a dry rough vacuum pump generate a chamber base pressure of a few $10^{-9}$ mbar and a few $10^{-8}$ mbar with and without cooling, respectively. In its current configuration, however, the system is not regularly baked because routine operation requires frequent replacement of substrates. Furthermore, although the ICA main chamber is separated from the projectile ion beamline (which operates at $10^{-6}$ mbar) by a series of gate valves and a differential pumping stage, it is likely that some 'pollution' from the beamline may be introduced during ice irradiation. However, since all operations in the experimental protocol (described fully in a later section) can be performed at a fixed sample position, and since relatively high current densities may be utilised, the effect of this pollution is expected to be negligible in most scenarios.

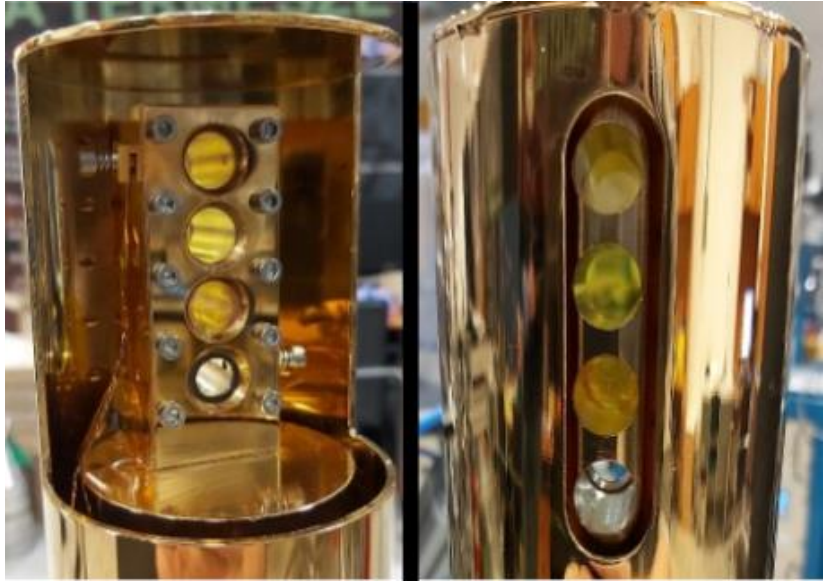

**Fig. 3** Photographs of the sample holder with the heat shield as seen from the front (left) and rear (right) sides. In these photographs, three zinc selenide substrates are mounted on the holder with the fourth substrate position being left vacant to serve as a 9.6 mm diameter collimator. The collimator, along with the Faraday cup $F_3$ pictured in Fig. 2, allows for ion beam lateral homogeneity determination. If the performed experiment is sensitive to beam homogeneity, then it is recommended that the arrangement presented here is used.

Ice analogues are prepared on the substrates via the background deposition of dosed gases or vapours. Pure and mixed gases are first introduced into a mixing container with the help of a mixer system of valves which is connected to gas cylinders or liquid-containing glass vials. Liquids are de-gassed in their vials through a series of freeze-thaw cycles. Pressure in the mixing container, which is typically a few mbar, is measured with a mass-independent capacitive manometer gauge (Edwards ASG2). The inlet system itself can also be carefully heated up and evacuated so as to reduce the risk of contamination.

Once the gases have equilibrated and are fully mixed in the mixing container, they are allowed into the main vacuum system through a fine regulating all-metal needle valve (LewVac). A distributor (scattering) plate mounted in front of the inlet tube reduces the pressure inhomogeneity in the chamber during gas injection so as to produce roughly the same ice thicknesses on all substrates (Table 1). Behind all four deposition substrates, the sample holder presents cylindrical holes of diameter 12 mm and length 30 mm in line with a wide slit opening in the heat shield to allow for FTIR spectroscopy (Fig. 3), while ensuring minimal gas deposition onto the rear side of the substrates (this is typically 4% of the total ice deposited on the front side; see the Appendix for more details on how this value is calculated).

**Table 1** Summary of experiments performed to demonstrate that ice thicknesses and compositions are similar across all four deposition substrates. Target thicknesses for the pure $CH_3OH$ ices were 0.5 μm, while for $CH_3OH$:$CO$ mixed ices (target composition of 40% $CH_3OH$), the target thicknesses were 0.75 μm. The techniques used to measure these parameters are explained in the next section. Although ice thicknesses are slightly lower at substrate 4 compared to the other substrates, this is usually not problematic as this substrate is generally replaced by a 9.6 mm collimator for ion beam lateral homogeneity determination as depicted in Fig. 3.

| Substrate | Pure $CH_3OH$ Ice Thickness (μm) | $CH_3OH$:$CO$ Mixed Ice Thickness (μm) | $CH_3OH$ Content of $CH_3OH$:$CO$ Mixed Ice |
|---|---|---|---|
| 1 | 0.47 | 0.74 | 36% |
| 2 | 0.51 | 0.77 | 38% |
| 3 | 0.49 | 0.76 | 39% |
| 4 | 0.41 | 0.64 | 41% |

The chamber itself is equipped with ten DN-40 CF ports separated by angles of 36° relative to each other which are used for external connections (Fig. 4). As is implied by Fig. 2, one of these ports serves as an entrance for the projectile ion beam delivered by the Tandetron accelerator, with a Faraday cup ($F_3$) mounted on the opposite port allowing for the beam quality to be monitored. The gas inlet system through which gases and vapours are introduced into the vacuum chamber for background deposition onto the substrates is located at the next port in the clockwise sense from this Faraday cup.

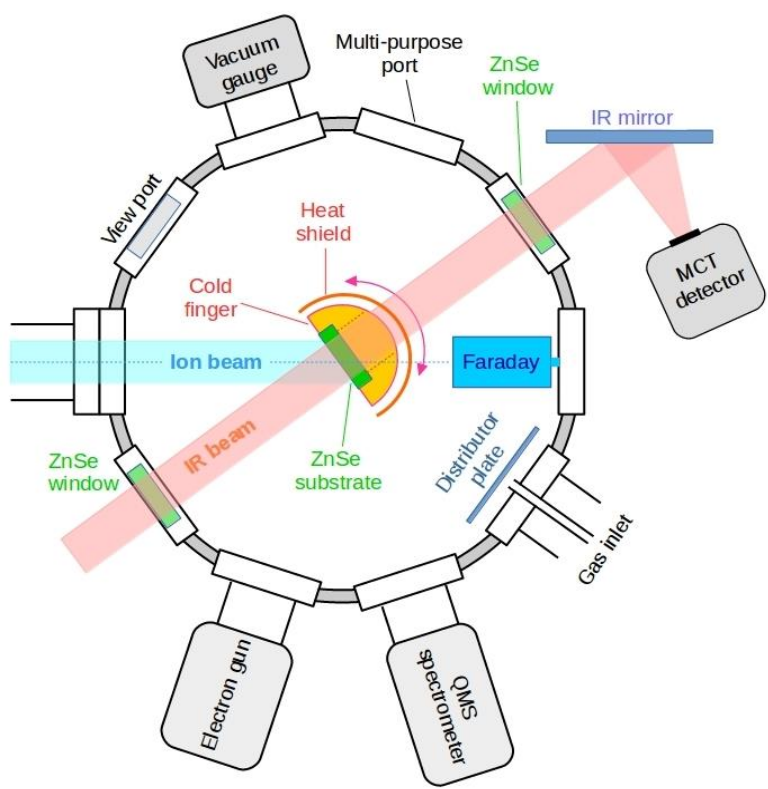

**Fig. 4** A top-view schematic diagram of the ICA chamber arranged for transmission FTIR absorbance spectroscopy. Although the sample holder and heat shield are rotatable, ion beam and electron beam irradiations are nominally performed as depicted, with the IR beam pathway orthogonal to the sample surface and charged projectiles impacting at angles of 36°.

The other ports on the chamber host an optical viewport which allows for direct observation of the sample holder and its substrates, a vacuum gauge (Pfeiffer Vacuum PBR260), a quadrupole mass spectrometer (QMS; Pfeiffer QME200), and an electron gun (Kimball Physics ELG-2A) operating over an electron energy range of 5-2000 eV. The characterisation of the emitted electron beams and the chemistry induced in astrophysical ice analogues as a result of irradiation by electrons is the subject of a separate publication [38].

Directly opposite to the electron gun is a spare multi-purpose port which, by default, is equipped with a Faraday cup which is used to monitor the electron beam current. The remaining two chamber ports are positioned opposite to one another (Fig. 4) and are fitted with zinc selenide windows so as to serve as the entrance and exit ports for the transmitted IR beam used in spectroscopic monitoring of the ice analogues.

## C. Analytical Techniques

The physico-chemical properties of pre- and post-processed astrophysical ice analogues are primarily monitored via FTIR absorbance spectroscopy nominally set up in transmission mode as shown in Fig. 4. Currently, a Thermo Nicolet Nexus 670 FTIR spectrophotometer with a spectral range of 4000-650 cm$^{-1}$ (2.5-15.4 μm) and a resolution of 1 cm$^{-1}$ is in use. In the present set-up (Fig. 4), the transmitted IR beam is detected by a mercury-cadmium-telluride (MCT) detector placed outside the main vacuum chamber in a small external container. A purge box continuously passes purified air through this MCT detector chamber, as well as the entire IR pathway, so as to prevent absorption by air constituents of variable composition (e.g., $H_2O$ or $CO_2$).

An important analytical capability of the set-up is that FTIR spectra can be taken continuously during deposition, irradiation, and thermal processing, thus helping to shorten the overall measurement time. Reflectance mode measurements are also possible by replacing the IR-transparent substrates with gold-coated mirror ones, and by using the port next to the ion beam entrance port in the clockwise direction to detect the reflected beam.

FTIR spectroscopy is not only used to qualitatively determine the nature of the molecular species present in the ice analogues via identification of their characteristic IR absorbance bands, but also to quantify their abundances. The column density $N$ (molecules cm$^{-2}$) of a given species can be calculated using measured peak areas for a characteristic absorbance band:

$$N = \frac{\int \tau(\nu)\,d\nu}{A_\nu} = \frac{P_{abs} \times \ln(10)}{A_\nu}$$

(Eq. 2)

where $\tau(\nu)$ is the optical depth of the ice component for a given wavenumber (cm$^{-1}$), $A_\nu$ is the integrated band strength for the measured IR band (cm molecule$^{-1}$), and $P_{abs}$ is the measured peak area of this absorbance band (cm$^{-1}$). Once calculated, the column density may be used to approximate the thickness $\theta$ (μm) of each component in the ice mixture using Eq. 3, with the estimated total ice thickness being the sum of all component thicknesses:

$$\theta = \left(\frac{mN}{\rho N_A}\right) \times 10^4$$

(Eq. 3)

where *m* and *ρ* are the respective known molecular mass (u) and density (g cm$^{-3}$) of the ice component in question, and $N_A$ is the Avogadro constant (6.022×10$^{23}$ mol$^{-1}$). It should be noted, however, that many values for $A_v$ quoted in the literature are valid for pure, single component ices and so using them to determine *N* for each molecular component in a mixed ice introduces an uncertainty which, in turn, influences the accuracy of the overall ice thickness estimation [39].

In light of this, we are in the process of installing a laser interferometer which will allow for a more accurate determination of the thicknesses of the deposited ice layers. The laser interference technique has been described previously [40] and is based on monitoring the variation in intensity of a He-Ne laser beam which is reflected off the surface of a substrate at a particular angle. During growth of the deposited astrophysical ice analogue, the laser intensity varies sinusoidally due to interference of the laser light reflected at the ice-vacuum and ice-substrate boundaries. The ice thickness is then calculated using Eq. 4:

$$\theta = \frac{\lambda_0 C}{2\eta \cos(\beta)}$$

(Eq. 4)

where $\lambda_0$ is the wavelength of the He-Ne laser light in a vacuum (0.6328 μm), *C* is the number of constructive pattern repetitions observed during deposition, *η* is the ice refractive index (which may be estimated from the ratio of the maxima and minima of the laser interference pattern), and *β* is the angle of the laser light in the ice.

The inclusion of a QMS attached to the main vacuum chamber (Fig. 4) also allows for a mostly qualitative assessment of the internal gas-phase composition. This is most relevant during ion radiolysis when components of the ice mixture may be sputtered as a result of ion impact [41], and during thermal processing during which solid-phase molecules may sublimate [42]. Presently, the QMS is able to detect sputtered and desorbed species with a molecular mass of up to 200 u with a mass resolution of 1 u, and is able to function efficiently at chamber pressures of up to ~10$^{-5}$ mbar.

**D. The Experimental Protocol**

In this section, we provide a description of the experimental protocol currently followed at the ICA during ion radiolysis studies of astrophysical ice analogues, assuming that the apparatus is set up for transmission mode FTIR absorbance spectroscopy. Firstly, the IR-transparent substrates (which, as previously mentioned, are typically zinc selenide) must be manually mounted onto the sample holder (Fig. 3). The zinc selenide deposition substrates typically used have a respective diameter and thickness of 15 mm and 3 mm.

Zinc selenide substrates have the option of being covered by a metallic mesh (Fig. 5), which allows for the avoidance of significant charge up and potential sparking at the sample surface when using comparatively high ion beam currents (>100 nA). The mesh is composed of a thin (5 nm) layer of chromium evaporated onto the surface of the substrate, on top of which is a thicker (250 nm) layer of gold. The width of these mesh lines is 20 μm, and their separation from each other is 0.8 mm.

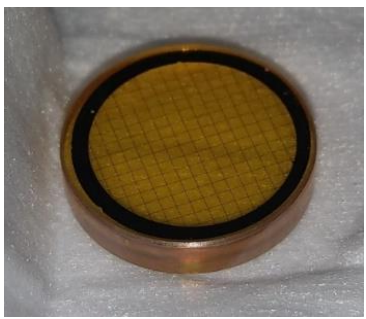

**Fig. 5** A close-up photograph of a zinc selenide substrate (diameter = 15 mm; thickness = 3 mm) with gold mesh evaporated onto its surface (mesh line width = 20 μm; thickness = 250 nm; separation = 0.8 mm). Between the gold mesh and the substrate is a thin chromium layer having a thickness of 5 nm.

Following insertion of the substrates, the ICA system is subsequently pumped down to a pressure of a few $10^{-8}$ mbar at room temperature. This is achieved by first engaging the dry vacuum pump until a chamber pressure of a few $10^{-2}$ mbar is reached, after which the turbomolecular pump is switched on for further pumping. The purge box supplying purified air to the MCT detector is also switched on at this point. Overall, the pumping down process requires several hours while the purge box requires in excess of one day to become fully operational.

Once a chamber pressure of a few $10^{-8}$ mbar is reached, the QMS is switched on and de-gassed. After the de-gassing (which typically takes a few minutes) is complete, the cryostat is switched on so as to cool down the sample holder from room temperature to the desired working temperature. For reference, cooling down to 20 K typically takes around two hours and at this temperature a chamber pressure of a few $10^{-9}$ mbar is recorded. Acquisition of background FTIR spectra of the zinc selenide substrates then follows.

The next phase in the experimental protocol is the deposition of the astrophysical ice analogues. Gases and vapours (the latter from de-gassed liquids) are introduced into the dosing line mixing chamber and allowed to equilibrate. The composition of the gas mixture is determined through the partial pressure of each component gas as read from the capacitive manometer gauge during its introduction. Once all components of the ice mixture have been introduced into the dosing line, the UHV all-metal fine needle valve leading to the main vacuum chamber is opened and the ices are formed via background deposition at a typical chamber pressure of ~$10^{-6}$ mbar.

During deposition, *in situ* FTIR spectra are collected so as to monitor the compositional stoichiometry and thickness of the growing ice using Eqs. 2 and 3, respectively. After the desired target ices have been successfully prepared, further background spectra of the deposited ice analogues may be collected.

Irradiation of the deposited ices is then performed using pre-defined fluence steps, with one or more FTIR spectra being collected after each step. The fluence $\phi$ (ions cm$^{-2}$) is defined as the number of ions impacting the ice per unit area integrated over the time of irradiation. Using the ICA, we are able to measure the total charge deposited at the sample during an irradiation step using a system of Faraday cups (Fig. 2) and can use this to calculate the corresponding fluence using Eq. 5:

$$\phi = \frac{\Delta Q}{Anq}$$

(Eq. 5)

where $\Delta Q$ is the total measured coulombic charge deposited at the sample, $A$ is the irradiated sample area (which in our system is 1.13 cm$^2$), $n$ is the integer charge state of the projectile ion, and $q$ is the fundamental charge constant (1.602×10$^{-19}$ C).

The fluence used during irradiation can be related to the energy imparted to the target ice as the projectile ion travels through it (i.e., the dose). The differential energy loss per unit path-length for an ion moving through a solid target is known as the stopping power and is the sum of elastic (nuclear stopping) and inelastic (electronic stopping) energy loss processes. It is most often calculated using a computer programme called *SRIM: The Stopping Range of Ions in Matter* [43]. The dose $D$ used during irradiation can be conveniently expressed in units of eV per equivalent small molecule (e.g., eV per 16 u), and is calculated as per Eq. 6:

$$D = \frac{\phi S_{\text{av}}}{\rho_{16}} \times 10^8$$

(Eq. 6)

where $S_{\text{av}}$ is the mean stopping power over the range of the projectile in the target ice (eV Å$^{-1}$) and $\rho_{16}$ is the molecular density of the hypothetical equivalent molecule of mass 16 u (molecules cm$^{-3}$).

The alternating pattern of irradiating using a pre-defined fluence step followed by FTIR spectral acquisition may be repeated until the desired final fluence (or dose) has been reached, with the QMS continuously detecting molecular fragments arising from sputtered or desorbed material. Finally, once irradiation has been concluded, thermal annealing and TPD of the processed ice analogues may be performed.

Although it is acknowledged that the apparatus is not currently designed to perform quantitative TPD experiments (especially in light of the background deposition presently used to create the astrophysical ice analogues), TPD may still be performed reliably after ion exposure of the deposited ice as long as special precautions are taken. In particular, only one substrate in the sample holder should be exposed to ion irradiation prior to TPD so as to exclude simultaneous desorption of product species from multiple sources (e.g., other substrates) during the TPD. In this scenario, kinetic FTIR spectroscopic monitoring of the single substrate during TPD studies provides reliable quantitative information on the ice, while qualitative gas-phase measurements made using the QMS should be corrected by means of selected control experiments to account for the sublimation of non-irradiated, less volatile icy material (e.g., $CH_3OH$ or $H_2O$) from other cold parts of the chamber. Re-condensation of sublimated material onto the irradiated deposition substrate is precluded due to homogeneous warming by the heating cartridge located within the sample holder, the geometry of the substrate holder which minimises deposition on the rear side of the zinc selenide substrates, and the presence of 77 K heat shield. We note that a directed deposition channel (wherein deposition is limited to single substrates) is presently under construction, and so a wider range of TPD studies can be performed in the near future.

## III. Preliminary Results

In this section, we present some experimental results from selected high-energy ion impacts of astrophysical ice analogues to show the potential use of the ICA end-station at Atomki. In particular, we report the results of amorphous $CH_3OH$ ice radiolysis by 200, 400, 750, and 1000 keV protons as well as 6 MeV $S^{2+}$ ions at 20 K under high-vacuum conditions. Our selection of $CH_3OH$ as a target ice was motivated by the fact that it is a molecule of astrochemical interest whose radiolytic processing is not only well described in the literature, but which may also lead to the formation of complex organic molecules [44-46].

For each experiment, the experimental protocol detailed above was followed: amorphous ice analogues of thickness 2.0±0.2 μm were created via the background deposition of the vapour onto zinc selenide substrates at 20 K. A pre-irradiation FTIR spectrum was recorded, after which the ices were individually processed by one of the selected high-energy ion beams. Additional spectra were collected at appropriate dose intervals so as to monitor the physico-chemical evolution of the ice.

### A. $CH_3OH$ Ice Processing by Energetic Protons

Continued irradiation of the $CH_3OH$ ice by energetic protons at energies of 200, 400, 750, and 1000 keV resulted in a decrease in the areas of the characteristic $CH_3OH$ absorption peaks (Fig. 6), while the radiolysis resulted in the concomitant appearance of new absorbance features attributed to the intra-molecular vibrational modes of radiolytic product species, including the molecules CO, $CO_2$, $CH_4$, and $H_2CO$, as well as the neutral radicals HCO and $CH_2OH$ (Fig. 6; Table 2).

The formation of these products is in good agreement with the results of previous irradiation studies of frozen $CH_3OH$ by protons [47,48], heavier ions [45,49], UV and X-ray photons [50-52], and electrons [44,46,53-55]. Comprehensive work by Schmidt *et al.* [54] and Bennett *et al.* [55] has elucidated much of the radiolytic reaction pathways available to $CH_3OH$, and thus the formation mechanisms of these product species have been characterised.

Initial fragmentation of $CH_3OH$ is thought to proceed by either loss of hydrogen to yield $CH_3O$ and $CH_2OH$ radicals, or alternatively via the loss of oxygen to yield $CH_4$. The formation of

CH$_4$ is also possible from the reaction between CH$_3$ radicals (which may themselves be produced directly from CH$_3$OH fragmentation) and a hydrogen atom abstracted from another species. Subsequent loss of hydrogen from CH$_3$O or CH$_2$OH affords H$_2$CO, which may then undergo a further loss of hydrogen to yield the HCO radical.

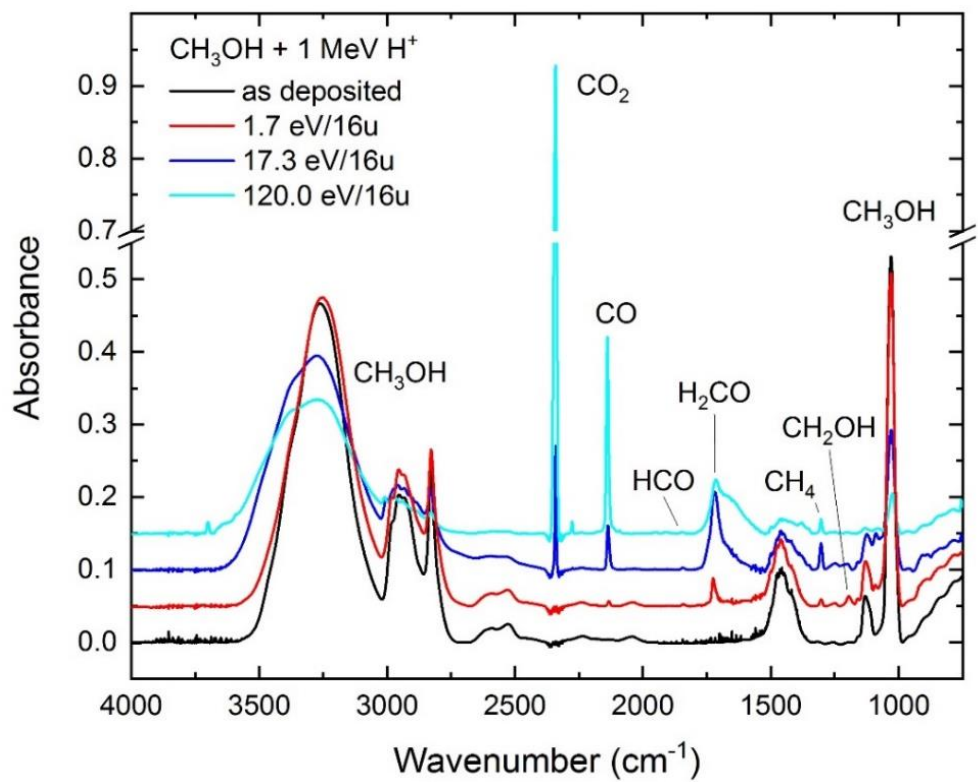

**Fig. 6** FTIR absorbance spectra of CH$_3$OH ice evolution under 1 MeV H$^+$ radiolysis. Note that individual spectra have been vertically offset for clarity.

**Table 2** List of species observed in the FTIR absorbance spectra of CH$_3$OH ice after radiolysis by 200, 400, 750, and 1000 keV protons and 6 MeV S$^{2+}$ ions. Products were identified via one of their characteristic absorbance bands, which are given in this table.

| Species | Absorbance Peak Used for Identification (cm$^{-1}$) | $A_v$ (10$^{-17}$ cm molecule$^{-1}$) | Reference |
|---|---|---|---|
| CH$_3$OH | 1020; $v_8$ | 1.61 | [56-58,62] |
| CO$_2$ | 2343; $v_3$ | 11.80 | [59,63] |
| CO | 2138; $v_1$ | 1.10 | [44,60] |
| H$_2$CO | 1725; $v_4$ | 0.96 | [44,64] |
| CH$_4$ | 1300; $v_4$ | 0.80 | [61,65] |
| HCO | 1843; $v_1$ | N/A | [25,44] |
| CH$_2$OH | 1195; $v_4$ | N/A | [25,44] |

The other products observed in Fig. 6, CO and CO$_2$, are considered to be second-generation products, meaning their formation is dependent upon the consumption of species which were initially produced by the radiolysis of $CH_3OH$. CO may be formed directly via the loss of hydrogen from either $H_2CO$ or HCO, with the former reaction being dominant [54,55]. Reaction of CO with OH then yields $CO_2$. A reaction network scheme for the reactions leading to the products observed in our experiments is given in Fig. 7.

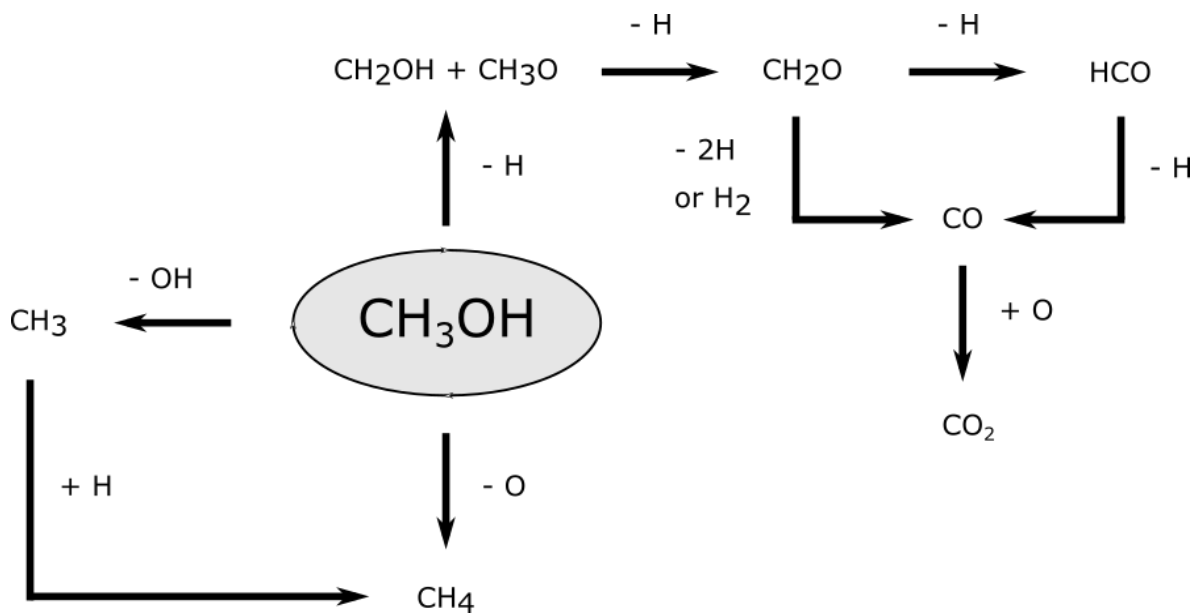

**Fig. 7** Reaction pathways leading to formation of products observed in the FTIR spectra after irradiation of frozen $CH_3OH$ by energetic protons and $S^{2+}$ ions [54,55]. These products are by no means the only species formed as a result of the irradiation of $CH_3OH$, and the formation of other, more complex organic molecules has also been reported in the literature [e.g., 46].

An interesting result of our study is that the exponential rate of decay in $CH_3OH$ molecular column densities with respect to dose was reproduced for all proton irradiations, irrespective of beam energy (Fig. 8). According to SRIM calculations, the range of the proton projectiles varied between 3-30 μm at the studied impact energies, meaning proton implantation into the ~2 μm ice is negligible. The similarity of the decay profiles seen in Fig. 8 may therefore be due to the fact that energy loss lies in the linear regime for all investigated proton irradiation experiments.

We have also investigated the variation of molecular column density with respect to dose for the product species $CH_4$. Our reasoning behind choosing this molecule for analysis was based on the fact that it possesses the most well-defined absorbance peak of all first-generation products in the measured FTIR spectra (Fig. 6). Our analysis (Fig. 8) shows that radiochemical $CH_4$ production peaks after a dose of ~30 eV per 16 u, after which it is depleted under prolonged irradiation. As was the case for the $CH_3OH$ decay profiles, it was noted that variations in $CH_4$ column density were well replicated for each proton beam energy used, likely for the same reasons.

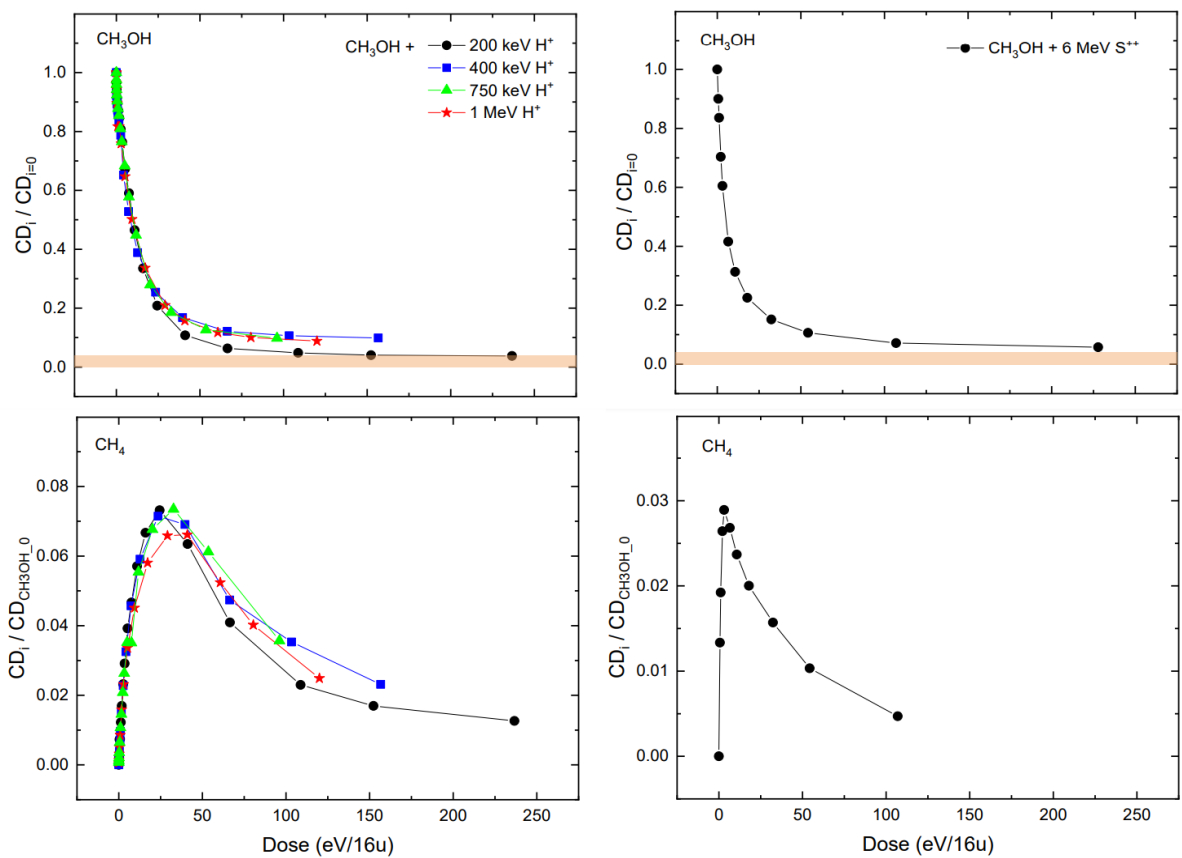

**Fig. 8** Top panels: Evolution of $CH_3OH$ column density as measured from the $\nu_8$ absorbance peak at ~1020 cm$^{-1}$ with increasing dose of proton (left) and $S^{2+}$ (right) ion beams. Bottom panels: Evolution of $CH_4$ column density as measured from the $\nu_4$ absorbance peak at ~1300 cm$^{-1}$ with increasing dose of proton (left) and $S^{2+}$ (right) ion beams. All column densities reported here are relative to the initial measured column density of $CH_3OH$. The pink horizontal bars in the top two panels indicate the lower bound value of $CH_3OH$ deposited on the rear side of the zinc selenide substrates (see Appendix for more details on how this is calculated).

### B. $CH_3OH$ Ice Processing by 6 MeV $S^{2+}$ Ions

The radiolysis of $CH_3OH$ ice induced by 6 MeV $S^{2+}$ ions was also investigated to provide a comparison with the results obtained in the proton beam experiments. In terms of the product species formed, no differences were noted between the two sets of experiments and all species recorded after proton irradiation of the ice (Fig. 6) were also recorded after $S^{2+}$ irradiation. It is expected, therefore, that the overall chemistry leading to the formation of these product molecules is similar to that outlined in Fig. 7. However, one noticeable difference between the two sets of experiments can be seen in the rate at which the $CH_3OH$ column density decays during irradiation (Fig. 8). This occurs significantly more rapidly during irradiation by 6 MeV $S^{2+}$ ions compared to irradiation by protons. Similarly, the radiochemical production of $CH_4$ peaks at a considerably lower dose before beginning to decay again.

It is difficult to pinpoint a single reason for this, as there are several possible factors which may contribute to this distinction in column density profiles. One possibility, for example, is that, although implantation is still negligible, the stopping power of the projectile $S^{2+}$ ion is not in the linear regime, and so results in a faster radiolysis of $CH_3OH$ and, consequently, a more rapid build-up and subsequent depletion of $CH_4$. Alternatively, it is also possible that due to the fact that the $S^{2+}$ ion is a much heavier projectile than a proton, the sputtering yield during this irradiation was significantly greater than that occurring during proton irradiation. This sputtering could lead to the removal of $CH_3OH$ and product molecules from the target ice [41], and thus also lead to a comparatively quicker depletion of these species. Indeed, QMS signals ascribed to $CH_3OH$ and stable molecular products were detected during this irradiation, indicating that ice sputtering occurred to at least some extent.

## IV. Concluding Remarks

In this paper, we have validated the capability of the newly commissioned ICA facility in Debrecen, Hungary to conduct ion radiolysis studies of interstellar ice analogues, and have also provided a detailed structural and operational description. Our preliminary studies irradiating amorphous $CH_3OH$ ice at 20 K under high-vacuum conditions are in agreement with prior work conducted using astrochemistry chambers at other facilities and show good repeatability and reproducibility.

Several planned upgrades will be carried out in the near future to expand the experimental capabilities of the ICA. For instance, an effusive evaporator is to be incorporated so as to allow for the deposition of more refractory materials such as complex organic molecules or ionic solids, including polycyclic aromatic hydrocarbons and amino acids. We also intend to increase our availability of spectroscopic monitoring methods by purchasing a UV-vis spectrophotometer, thus extending the possible observable spectral range.

External access to the ICA is presently managed through the Europlanet 2024 Research Infrastructure: a European Union funded project under the Horizon 2020 Programme, of which the ICA is one of several constituent transnational access (TA) facilities. Project applications may be submitted by research groups from around the world in response to TA calls, which currently take place annually. Successful applicants may then perform their study either as an in-person visit by working in collaboration with the local research team, or remotely through a virtual access procedure.

## Acknowledgements

The authors all gratefully acknowledge support from the Europlanet 2024 RI which has received funding from the European Union's Horizon 2020 Research Innovation Programme under Grant Agreement No. 871149. The main components of the ICA set-up were purchased with funds from the Royal Society through grants UF130409, RGF/EA/180306, and URF/R/191018. This work has also been supported by the European Union and the State of Hungary; co-financed by the European Regional Development Fund through grant GINOP-2.3.3-15-2016-00005. Support has also been received from the Hungarian Scientific Research Fund (Grant No. K128621). We would also like to extend our thanks to the technical and operational staff at the Tandetron Laboratory in Debrecen, Hungary for their assistance.

Duncan V. Mifsud is the grateful recipient of a University of Kent Vice-Chancellor's Research Scholarship. Sergio Ioppolo acknowledges the Royal Society for financial support. The research of Zuzana Kaňuchová is supported by VEGA – the Slovak Grant Agency for Science (Grant No. 2/0023/18) and the Slovak Research and Development Agency (Contract No. APVV-19-0072). Alejandra Traspas-Muiña thanks Queen Mary University of London for funding. The research of Béla Paripás is supported by the European Union and the Hungarian State; co-financed by the European Regional Development Fund (Grant GINOP-2.3.4-15-2016-00004).

## Appendix: Calculation of the Amount of Ice Deposited on the Rear Side of the Substrate During Background Deposition

The sticking of gas-phase molecules to cold surfaces is a complex physical phenomenon. At the microscopic level, the probability of a successful sticking event is dependent upon the nature of the incident molecule, its orientation, energy, and angle of incidence; as well as macroscopic parameters, such as the material, structure, and temperature of the cold surface. Moreover, non-sticking is not solely a single scattering event, and the energy and angular distribution of reflected molecules may be of significance. At the macroscopic level, the gas-phase molecular material may be characterised by its pressure and temperature. For a given set of values of the macroscopic parameters, a sticking coefficient $c_s$ (i.e., the ratio of the number of successful sticking events to the number of collisions between gas-phase molecules and the cold surface) can be defined.

In our calculations we assume that, during deposition of gas-phase molecules onto the surface of a cold ice composed of the same molecules, such a sticking coefficient is large. Indeed, at 20 K we consider it to be close to unity. Once the first few molecular monolayers of a homogeneous, single component gas have been deposited, then the material compositions of the gas-phase and ice-phase are identical. This may not necessarily be the case for mixed, multi-component gases, but the similarities between the gas and ice phases may be determined experimentally by controlling the stoichiometric ratio of the components of the inlet gas and by measuring this same ratio for the deposited ice using FTIR spectroscopy.

In order to model the ratio of the thickness of the ices deposited on the front and rear sides of a substrate, we assume a sticking coefficient of unity as a first approach. Thus, if a molecule touches a cold surface, then we consider it to successfully stick to that surface. Furthermore, at deposition pressures of <$10^{-4}$ mbar, molecules only interact with the surfaces of the chamber, and do not collide with each other. As such, the only molecules which successfully stick to the substrate are those which make contact with it having not collided with any other cold surface along their way. Accordingly, the deposition yield at a point $r$ on the substrate is proportional to the acceptance solid angle $\Omega(r)$, which is in turn determined by the geometry of the cold sample holder. The effective deposited thickness is thus proportional to the average solid angle, given by Eq. A1:

$$\bar{\Omega} = \frac{1}{A}\int_A d\vec{r}\,\Omega(\vec{r})$$

(Eq. A1)

where $A$ is the relevant surface area illuminated by the FTIR spectroscopic beam.

This average solid angle is significantly different for the front and rear sides of the substrate. The geometrical arrangement of our set-up is shown in Fig. 9. Calculating the acceptance solid angles is a lengthy but straightforward process. As a result, the effective relative thickness $d_{rel}$ of the ice deposited on the rear side of the substrate (assuming $c_s = 1$) is given as:

$$d_{rel} = \frac{d_{rear}}{d_{rear} + d_{front}} = \frac{\bar{\Omega}_{rear}}{\bar{\Omega}_{rear} + \bar{\Omega}_{front}} = 0.039$$

(Eq. A2)

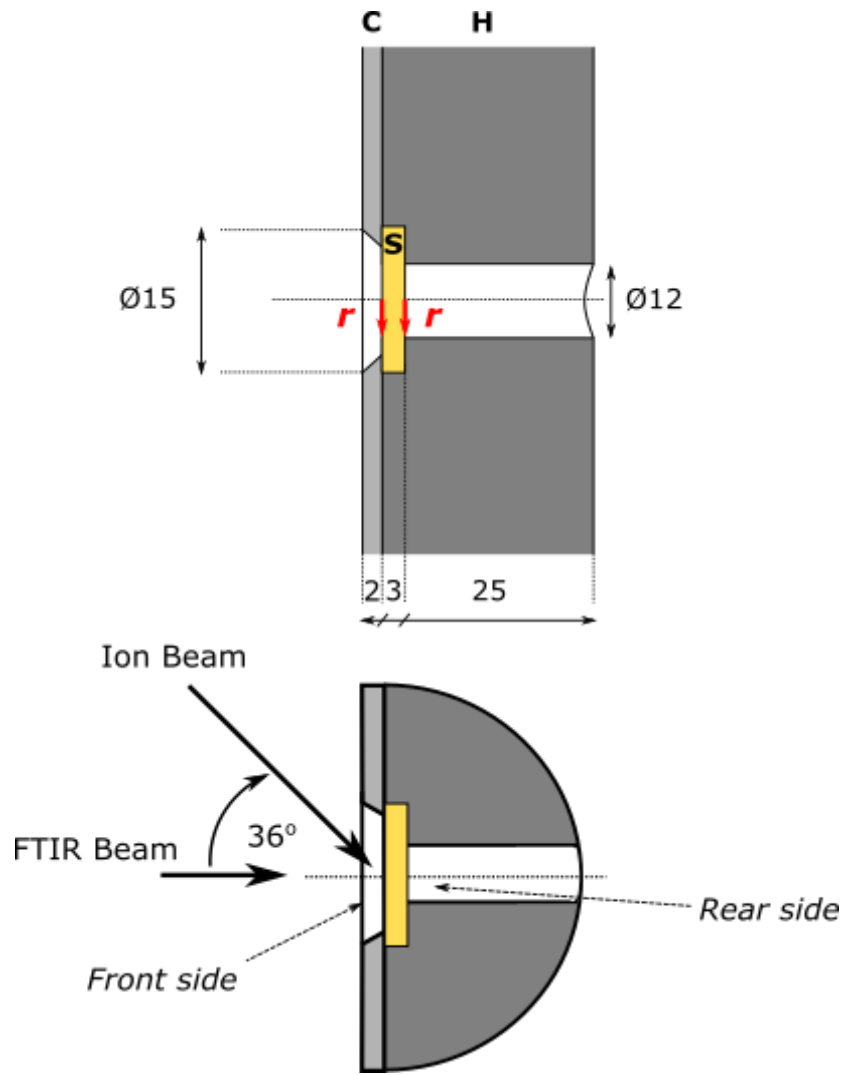

**Fig. 9** The geometry of the gold-coated sample holder H, clump plate C, and zinc selenide substrates S. All parts are cooled to the same temperature. Length values presented in this figure are given in mm.

Note that this relative thickness may be fairly accurately approximated by the ratio of the central ($r = 0$) solid angles $\Omega°$. For the rear side, this is calculated using the average length of the hole within the copper sample holder; 24.7 mm.

$$\Omega°_{rear} = 2\pi(1 - \cos(\alpha_{rear})) = 2\pi\left(1 - \frac{24.7}{\sqrt{24.7^2 + 6^2}}\right) = 0.178$$

(Eq. A3)

$$\Omega°_{front} = 2\pi(1 - \cos(\alpha_{front})) = 2\pi\left(1 - \frac{2}{\sqrt{2^2 + 7.5^2}}\right) = 4.664$$

(Eq. A4)

$$d_{rel} = \frac{\Omega°_{rear}}{\Omega°_{rear} + \Omega°_{front}} = \frac{0.178}{0.178 + 4.664} = 0.037$$

(Eq. A5)

where $\alpha$ is the half-angle of the relevant solid angle cone.

These calculations imply that only around 4% of the total thickness monitored using FTIR spectroscopy remains unirradiated. It should be noted, however, that this is the lower limit value for $d_{rel}$, which may be larger if the sticking coefficient is in fact smaller. For instance, if $c_s = 0.5$ (i.e., only half of collisions with the rear side result in successful sticking) a numerical simulation based on the assumption that the other half of collisions result in specular reflection yielded a relative thickness ($d_{rel}$) of to 0.094, or 9.4%.

Therefore, we conclude that in most cases where the deposition rate is large (i.e., $c_s > 0.5$), the thickness of the non-irradiated layer on the rear side of the substrate is <10% of the total deposited thickness. In the 20 K irradiations presented in this study, the average asymptotic value of the intact (i.e., not destroyed) $CH_3OH$ ice component was 7.25% with a scatter of ±2.50%, which is in reasonable agreement with our approximation.

## Data Availability

The data that support the findings of this study are available from the corresponding author upon reasonable request.